\documentclass{PoS}

\title{Leading-order hadronic contributions to $g_\mu-2$}

\ShortTitle{Hadronic contributions to $g_\mu-2$}

\author{\begin{minipage}{\textwidth}
\begin{center}
\textbf{ Budapest-Marseille-Wuppertal Collaboration}\\[0.2cm]
\speaker{Eric B.~Gregory}$^{1}$, Zoltan Fodor$^{1,2,3}$, Christian Hoelbling$^1$,
Stefan Krieg$^{1,2}$,
Laurent Lellouch$^{4,5}$, Rehan Malak$^{4,5,6}$,  Craig McNeile$^1$, Kalman Szabo$^7$\\[0.4cm]
{\it
$^1$Department of Physics, Bergische Universit\"{a}t Wuppertal, Gaussstr. 20, 
D-42119 Wuppertal, Germany\\
$^2$IAS/JSC, Forschungszentrum J\"ulich, D-52425 J\"ulich, Germany\\
$^3$Inst. for Theor. Physics, E\"otv\"os University, P\'azm\'any P. s\'et. 1/A, H-1117
Budapest, Hungary\\
$^4$ Aix-Marseille Universit\'e, CNRS, CPT, UMR 7332, F-13288 Marseille,
France\\
$^5$ Universit\'e de Toulon, CNRS, CPT, UMR 7332, F-83957 La Garde,
France\\
$^6$ CNRS, CEA, Maison de la Simulation,
USR 3441, F-91191 Gif-sur-Yvette Cedex, France\\
$^7$University of Regensburg, Universit\"atsstr. 31, 93053 Regensburg}
\end{center}

 E-mail: \email{gregory@uni-wuppertal.de}
\end{minipage}

}

\abstract{We present preliminary lattice results for the leading-order 
hadronic contribution to the muon anomalous magnetic moment, calculated with 
HEX-smeared clover fermions. In our calculation we include 2+1-flavor ensembles 
with pions at the physical mass.}

\FullConference{31st International Symposium on Lattice Field Theory - LATTICE 2013\\
		July 29 - August 3, 2013\\
		Mainz, Germany}

\newcommand{\mr}{\mathrm}

\newcommand{\Mpi}{M_\pi}

\begin{document}
\vspace{-0.15in}
\section{Introduction}
The muon anomalous magnetic moment is among the most precisely 
measured quantities in physics with $a_\mu\equiv\frac{(g_\mu-2)}{2}$
determined experimentally to about 0.5 parts per million
\cite{Bennett:2006fi}.
Theoretical calculations of Standard Model contributions to $a_\mu$ have similar
precision. There currently exists tension between Standard Model and 
experimental determinations of 3.6 standard deviations \cite{PGDmug-2}:
\begin{equation}
a_\mu^{\rm exp} -a_\mu^{\rm SM}  = 287(63)(49)\times 10^{-11}.
\end{equation}
The possibility that this tension is a hint of beyond Standard Model physics
has led to renewed effort to improve the precision of these determinations.
The Muon $g-2$ experiment at Fermilab aims to improve the experimental 
precision to 0.14 parts per million \cite{FNALweb}.

The full standard model calculation includes contributions from QED,
electro-weak and hadronic processes.
The uncertainty on the theory side is dominated by the calculation of the
hadronic contributions. The current best precision of the leading such 
contribution, known as the hadronic vacuum polarization (HVP) contribution, 
comes from experimental 
$e^+e^-$ cross-section data 
\cite{ee-tau-Davier,ee-tau-Hagiwara}   
and $\tau \rightarrow \nu_\tau +$ {\it hadrons} decay data \cite{taudecay}.

The challenge is for lattice QCD to provide first-principle calculations of 
the hadronic contributions to $a_\mu$ that meet or exceed the current 
precision of semi-empirical methods. 
There have been a number of attempts by different lattice groups 
5\cite{Aubin:2006xv, Boyle:2011hu, Feng:2011ff,Gockeler:2003cw,DellaMorte:2012cf}
demonstrating the feasibility of the approach.
A full calculation will require a calculation of the hadronic vacuum 
polarization (HVP), including disconnected contributions, as well as the
contribution of light-by-light scattering through hadrons.

Here we give a preliminary report of our efforts to calculate the leading-order
contribution of the HVP. We present results based on
lattices with either 2+1 flavors or four non-degenerate flavors of HEX-smeared
clover-type fermions. We include ensembles with pion masses at or below the 
physical value.
\vspace{-0.1in}
\section{Lattice calculation}
Our preliminary calculations have been performed on the ensembles listed in 
Table \ref{tab:cfgs}. We use HEX-smeared clover-type fermions. We use either
two or three levels of HEX smearing. The ``2-HEX'' ensembles have $N_f=2+1$ 
flavors and are described more fully in \cite{Durr:2010aw}. These include 
ensembles with the pion mass at or below the physical value. The ``3-HEX''
lattices
have four non-degenerate flavors of dynamical fermions, corresponding to
$u$, $d$, $s$ and $c$ quarks.

\begin{figure}
\begin{center}
\includegraphics[width=.2\textwidth]{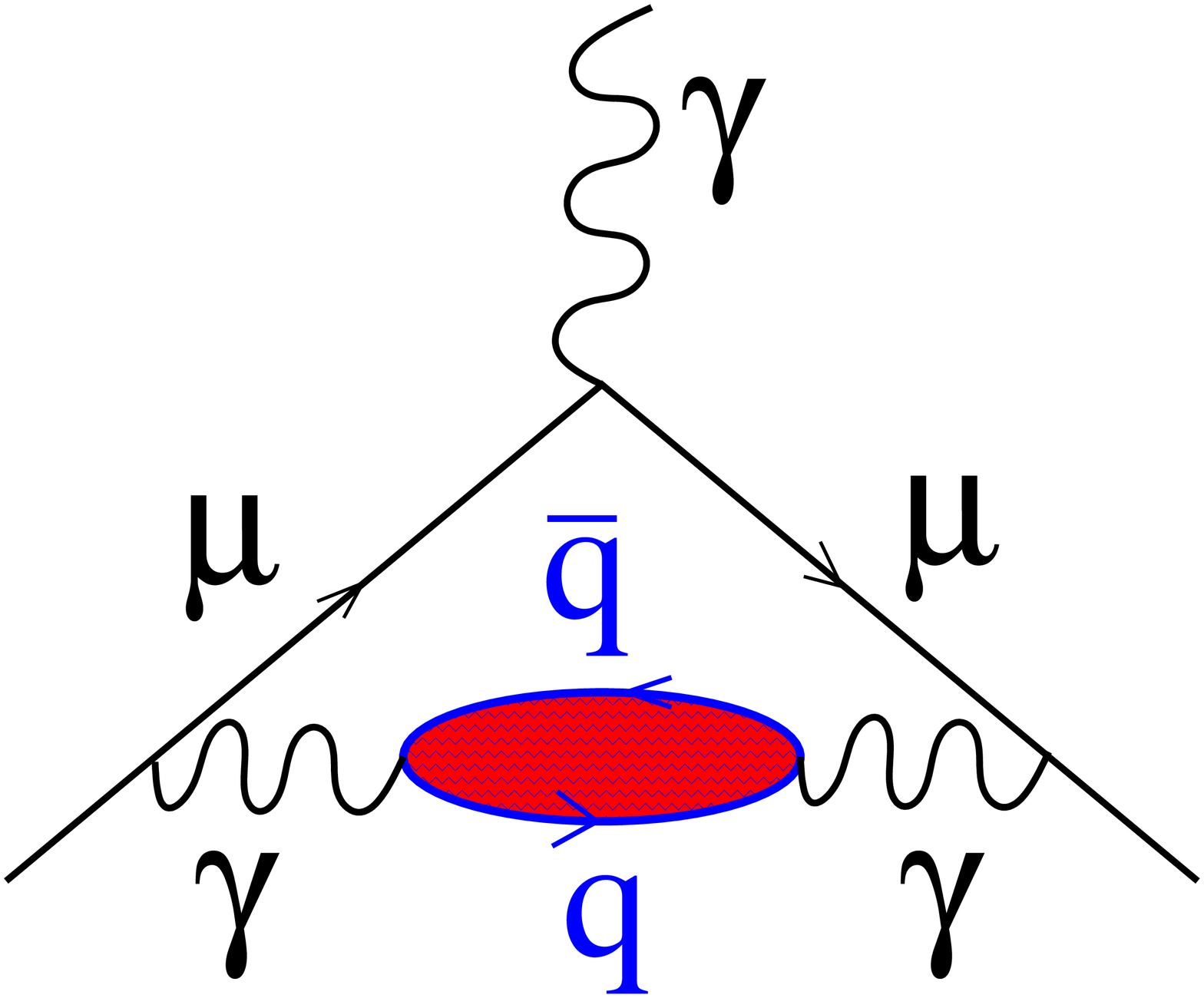}
%\hfill
\hspace{1in}
\includegraphics[width=.2\textwidth]{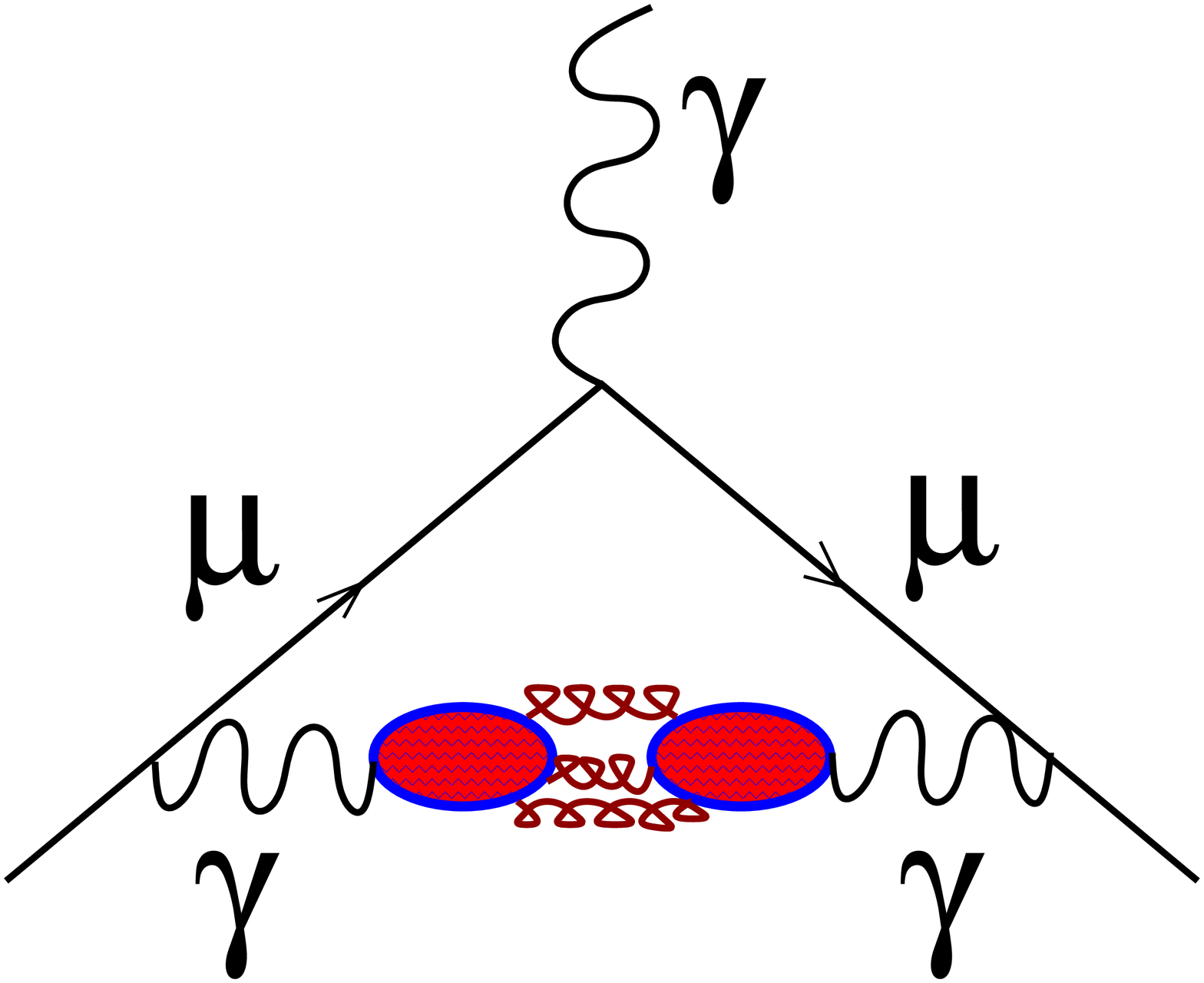}
\end{center}
\caption{Leading-order connected and disconnected hadronic contributions 
to $a_\mu$.}
\label{HVP_lead}
\end{figure}

\begin{table} 
{\small
\begin{center}
%\begin{tabular}{cccccc}
\vspace{-0.1in}
\begin{tabular*}{\textwidth}{c @{\extracolsep{\fill}}ccccc}
\hline
\hline
\multicolumn{6}{c}{{\bf 2-HEX ($N_f = 2+1$)}}\\
$am_{ud}^\mr{bare}$ & $am_s^\mr{bare}$ & volume & \#\,cfgs\ & $\Mpi$ (GeV) &  $n
_{\rm tw}$\\

\hline
\multicolumn{6}{c}{$\beta = 3.31$, $a^{-1} = 1.697$ GeV}\\
-0.09933 & -0.0400 & $48^3\times 48$ & 928 & 0.136(2)\\
-0.09300 & -0.0400 & $24^3\times 48$ & 210 & 0.255(2)\\
\hline
\multicolumn{6}{c}{$\beta = 3.5$, $a^{-1} = 2.131$ GeV}\\
  -0.05294 & -0.0060 & $64^3\times 64$ & 83  & 0.130(2)\\
  -0.04900 & -0.0120 & $32^3\times 64$ & 216 & 0.250(2)\\
  -0.04900 & -0.0060 & $32^3\times 64$ & 110 & 0.258(2)\\
  -0.04630 & -0.0120 & $32^3\times 64$ & 212 & 0.308(2)\\
\hline
\multicolumn{6}{c}{$\beta = 3.61$, $a^{-1} = 2.561$ GeV}\\
 -0.03000 & -0.0042 & $32^3\times 48$ & 188 & 0.332(4) & 0.5, 0.25, 0.1\\
\hline
\multicolumn{6}{c}{$\beta = 3.7$, $a^{-1} = 3.026$ GeV}\\
 -0.02700 &  0.0000 & $64^3\times 64$ & 208 & 0.182(2)\\
%\hline
\end{tabular*}\\
%\end{center}
%\begin{center}
\vspace{0.1in}
%\begin{tabular}{ccccccc}
\begin{tabular*}{\textwidth}{c @{\extracolsep{\fill}}cccccc}
\hline
\hline
\multicolumn{7}{c}{{\bf 3-HEX ($N_f = 4$)}}\\
 $am_{u}^\mr{bare}$ &  $am_{d}^\mr{bare}$ & $am_s^\mr{bare}$ & $am_c^\mr{bare}$ &volume & \#\,cfgs\ & $\Mpi$ (GeV) \\
\hline

\multicolumn{7}{c}{$\beta = 3.2$, $a^{-1} = 1.897$ GeV}\\
-0.0806 & -0.0794 & -0.033 & 0.71 &  $32^3\times 64$ & 240 & 0.250\\
\hline
\hline
\end{tabular*}
\end{center}
}
\caption{\label{tab:cfgs}Configurations used in preliminary study.}
\end{table}
The contribution of the HVP at the lowest order 
comes from diagrams such as Fig.~\ref{HVP_lead}. The lattice method devised by 
Blum \cite{Blum:2002ii} is based on the recognition that these diagrams 
can be calculated
by determining the vacuum-subtracted HVP, $\hat{\Pi}(Q^2)$ as a function 
of the square of the Euclidean momentum $Q$, then integrating \cite{Lautrup:1971jf}%\cite{Lautrup:1971yp}
\begin{equation}
\label{HVP_integral}
a_\mu^{\rm had,LO} = \frac{\alpha}{\pi}\int^\infty_0dQ^2f(Q^2)\hat{\Pi}(Q^2),
\end{equation}
 with the kernel function
\begin{equation}
f(Q^2) = \frac{m^2_\mu Q^2Z(Q^2)^3 \left(1 - Q^2Z(Q^2)\right)}
{1 + m_\mu^2Q^2Z(Q^2)^2},
\end{equation}
where
\begin{equation}
Z=-\frac{Q^2 - \sqrt{Q^4 + 4m^2_\mu Q^2}}{2m_\mu^2Q^2}.
\end{equation}

On the lattice we calculate for each flavor, $f$, the HVP tensor as the Fourier 
transform of the vector current correlator:
\begin{equation}
\Pi_{\mu\nu}^f(\hat{Q}) = a^4 \sum_y e^{iQ(x+\frac{a\hat{\mu}}{2}-y)}\langle J_\mu^{\rm CVC}(x)J_\nu^{\rm loc}(y)\rangle
,
\end{equation}
with 
\begin{equation}
 J^{{\rm loc},f}_\nu(y) = \overline{\psi}^f(x)\gamma_\nu\psi^f(x),
\end{equation}
and the conserved vector current (CVC) as given by
\begin{equation}
J_{\mu}^{{\rm CVC},f}(x)= \frac{1}{2}  \Big[\bar{\psi}^f (x+a \hat{\mu}) 
(1+\gamma_{\mu}) U_{\mu}^{\dagger}(x) \psi^f (x) 
 -\bar{\psi}^f (x) (1-\gamma_{\mu}) U_{\mu}(x) \psi^f (x+a \hat{\mu}) \Big]
.
\label{curr1}
\end{equation}
The HVP tensor satisfies the Ward-Takahashi Identity (WTI) on the conserved index $\mu$: 
\begin{equation}
\hat{Q}_\mu\hat{\Pi}_{\mu\nu}^f = 0,
\end{equation}
with the modified lattice momentum
\begin{eqnarray}
\hat{Q}_\mu = \frac{2}{a}\sin\left(\frac{aQ_\mu}{2}\right)&\,\,\,\,{\rm and}\,\,\,\,&
Q_\mu = \frac{2\pi n_\mu}{L_\mu}.
\end{eqnarray}
To enforce conservation on the local current sink index $\nu$ we require 
$Q_\nu=0$. We also use diagonal $\mu=\nu$ elements only.

With Euclidean momentum $Q_\mu$, the vacuum-subtracted HVP scalar $\hat{\Pi}(Q^2)$ appearing in 
(\ref{HVP_integral}) is 
related to the HVP tensor $\Pi_{\mu\nu}^f(Q)$ through
\begin{equation}
\Pi_{\mu\nu}^f(Q) = \left(Q^2\delta_{\mu\nu} -Q_{\mu}Q_{\nu}\right)\Pi^f(Q^2)
\end{equation}
and
\begin{equation}
\label{subtraction}
\hat{\Pi}(Q^2) = 4\pi\alpha\sum_{f=0}^{N_f}q_f^2\left(\Pi^f(Q^2) - \Pi^f(0)\right),
\end{equation}
where $q_f$ is the electromagnetic charge of quark flavor $f$. 

To perform the vacuum subtraction in (\ref{subtraction}) we must know the value 
of $\Pi^f(0)$, which is not directly accessible from the lattice data.
To do so we fit the measured 
values of $\Pi^f(0)$ to a suitable function of $Q^2$ and extrapolate to $Q^2 = 0$.
For simplicity in this preliminary work we fit to:
\begin{equation}
\Pi(Q^2) = c + \sum_{i=0}^N\frac{ b_i}{Q^2 + c_i},
\label{fit_form}
\end{equation} 
a multi-vector-dominance model, with $N=1$ or 2, as the data support.
Golterman {\it et al.}\cite{Golterman:2013vca} note 
that this is not an optimal fit ansatz. In the final calculation we will 
explore different fit forms to constrain systematic errors.

\begin{figure}[h]
\begin{center}
\includegraphics[width=0.35\textwidth]{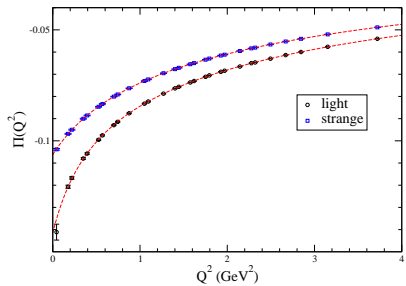}
\end{center}
\caption{Sample fit of light and strange components of the HVP scalar from 
$\beta = 3.50$ $M_\pi = 250$ MeV data set.
\label{samp_fit}}
\end{figure}

\begin{figure}[h]
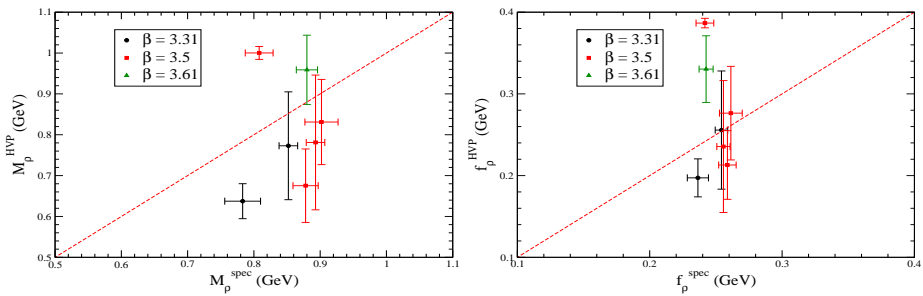

\vspace{-0.1in}
\begin{center}
\includegraphics[width=0.40\textwidth,height=1.5in]{FIGS/fit_param_plot1.eps}
\includegraphics[width=0.40\textwidth,height=1.5in]{FIGS/fit_param_plot2.eps}
\end{center}
\vspace{-0.25in}
\caption{
Comparison of $M_\rho$ and $f_\rho$ from HVP fits and spectroscopy fits.
\label{sane_check}}
\end{figure}

An example of the fits to unsubtracted HVP scalars is shown in Fig.~\ref{samp_fit}. The vector dominance model suggests that the HVP scalar should behave
approximately as
\begin{equation}
\Pi^{\rm tree}(Q^2) = \frac{2}{3}\frac{f^2_V}{Q^2 + m^2_V}.
\end{equation}
As a consistency check we compare values of $M_\rho$ and $f_\rho$
obtained from the fits ($M_\rho^{\rm HVP} \equiv c_0^{1/2}$ and 
$f_\rho^{\rm HVP} \equiv \sqrt{3b_0/2}$) respectively 
with those extracted from straightforward 
spectroscopy fits of the zero-momentum correlators. These comparisons are shown in Figs.~\ref{sane_check}a and \ref{sane_check}b.

To determine $a_\mu^{\rm had,LO}$ we use the fitted parameters to define a continuous
function $\Pi(Q^2)$ with (\ref{fit_form}), substitute the resulting 
$\hat{\Pi}(Q^2)$ into the integral (\ref{HVP_integral}), which we evaluate 
numerically.

\subsection{Twisted boundary conditions}
The integrand $f(Q^2)\hat{\Pi}(Q^2)$ has a peak at around the muon mass, which 
is approximately an order of magnitude lower than the smallest, non-vanishing
lattice momentum available on our lattices.  This creates a 
large model-dependence as we extrapolate our results toward $Q^2=0$.

Twisted boundary conditions have been proposed \cite{Sachrajda:2004mi} 
as a method of accessing arbitrarily low lattice momenta.
One must
twist the spatial boundary conditions in the valence quark and anti-quark
fields by a relative angle 
\begin{equation}
\psi(x+L_\mu \hat{\mu}) = e^{i\theta_\mu^{\rm tw}}\psi(x)
\,\,\,\,\,\,\,\,{\rm with}\,\,\,\,\,\,\,\,\,\,
\theta^{\rm tw}_\mu = 2\pi n^{\rm tw}_\mu.
\end{equation}
The lattice momenta transform as 
\begin{equation}
Q_\mu \rightarrow Q_\mu - \theta_\mu^{\rm tw}/L_\mu
\end{equation}
in the twisted direction(s).

\begin{figure}[h]
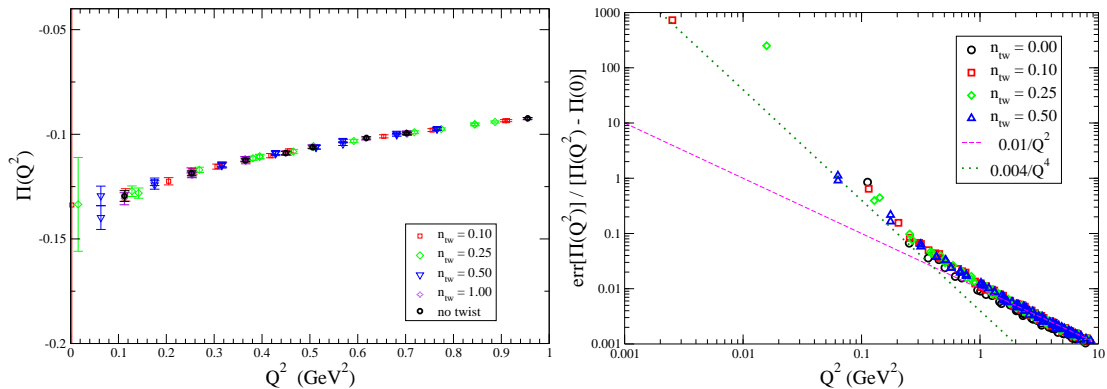

\vspace{-0.1in}
\begin{center}
\includegraphics[width=0.48\textwidth]{FIGS/hex2_b3.61_m0.0300_m0.0042_32_48_m0.0_twist_scalar.eps}
\includegraphics[width=0.48\textwidth]{FIGS/twistedBC_errors_sig_hx2_b3.61_m0.03_m0.0042_m0.0.eps}
\end{center}
\vspace{-0.25in}
\caption{(left) Comparison of twisted BC and non-twisted BC data for 
the light quark channel of the $\beta=3.61$ $M_\pi = 332$ MeV ensemble. 
(right) Error/signal for the same points. Dashed lines to guide the eye.
\label{twBC_data}}
\end{figure}

We explore this (Fig. \ref{twBC_data}) and note several issues. 
First, the naive twisting breaks the WTI, though the violation becomes 
negligible as the spatial volume increases. Aubin {\it et al.} 
\cite{Aubin:2013daa} note this and provide a term to correct it. 
Second, the relative statistical error on $\hat{\Pi}(Q^2)$ grows approximately 
like $1/Q^4$ at low $Q^2$ due to the division by 
$\left(Q^2\delta_{\mu\nu} -Q_{\mu}Q_{\nu}\right)$
and the subsequent subtraction of the $Q^2=0$ value. At our current
statistics,  the new twisted points serve mainly as a consistency check
without constraining the fit function significantly. 
We have not included twisted BC data in the preliminary results in the next section.
\begin{figure}[h]
\begin{center}
\includegraphics[width=0.40\textwidth]{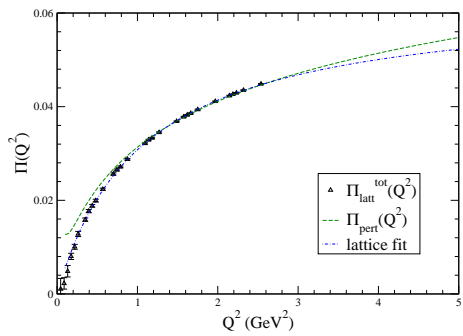}
\end{center}
\vspace{-0.25in}
\caption{
A sample matching of the lattice data to perturbation theory for $\beta=3.5$,  $M_\pi=130$ MeV.
\label{fig:pertmatch}}
\end{figure}

\subsection{Matching to perturbation theory}
A careful calculation of $a_\mu^{\rm had,LO}$ should include a matching of lattice data
to perturbation theory at large values of $Q^2$. 
In Fig. \ref{twBC_data}b we demonstrate 
that such matching is feasible for the $Q^2\approx 2$ GeV region, 
using expressions from \cite{Chetyrkin:1996cf}.
We do not include such a matching in our current calculation, introducing 
systematic error of $\sim 1\%$ or less.
\vspace{-0.15in}
\section{Results and conclusions}
\begin{figure}[t]
\vspace{-1.5in}
\begin{center}
\includegraphics[width=.45\textwidth, height = 0.64\textwidth]{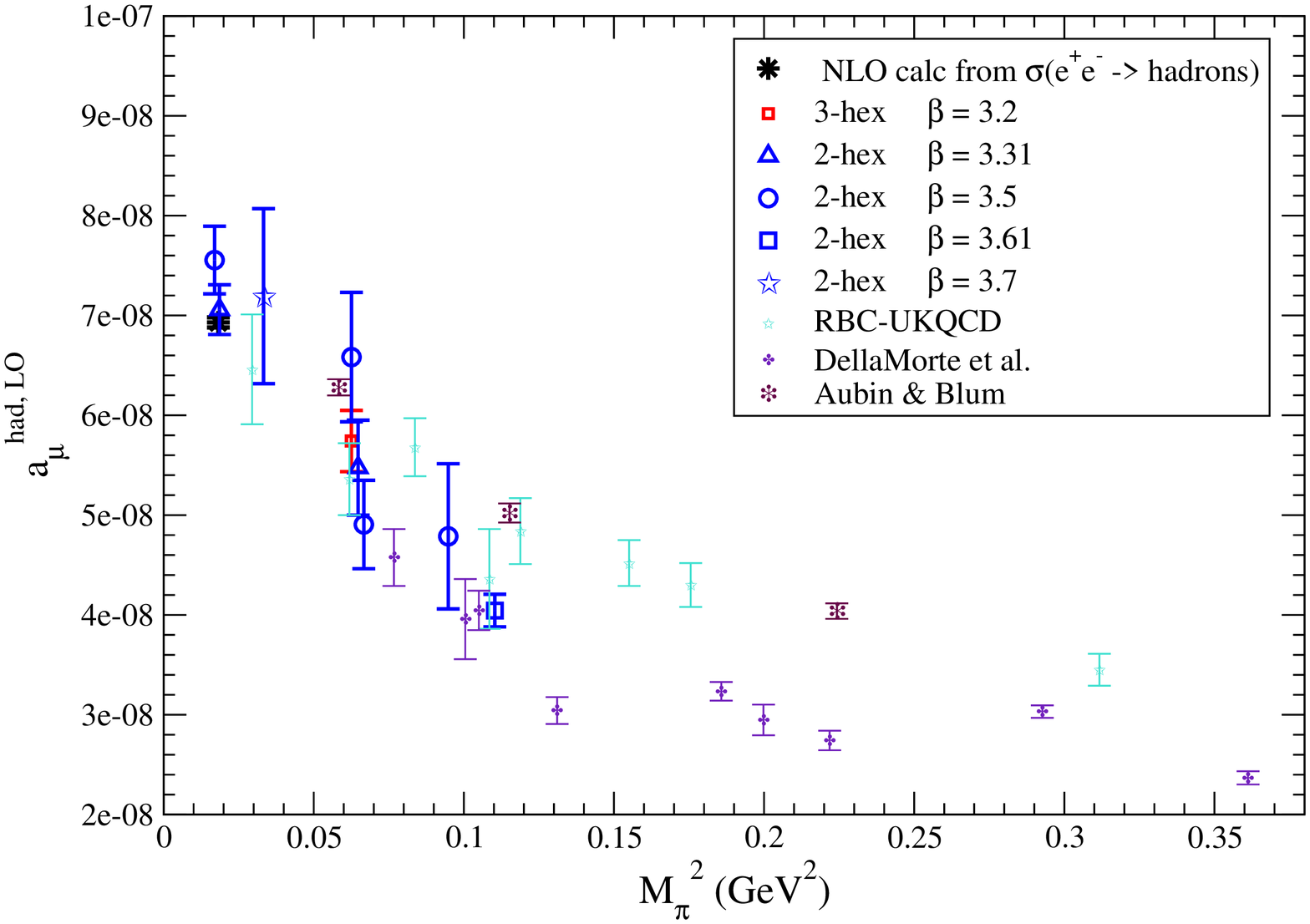}
\hspace{0.35in}
\includegraphics[scale=.43,trim=74 150 73 220,clip=true]{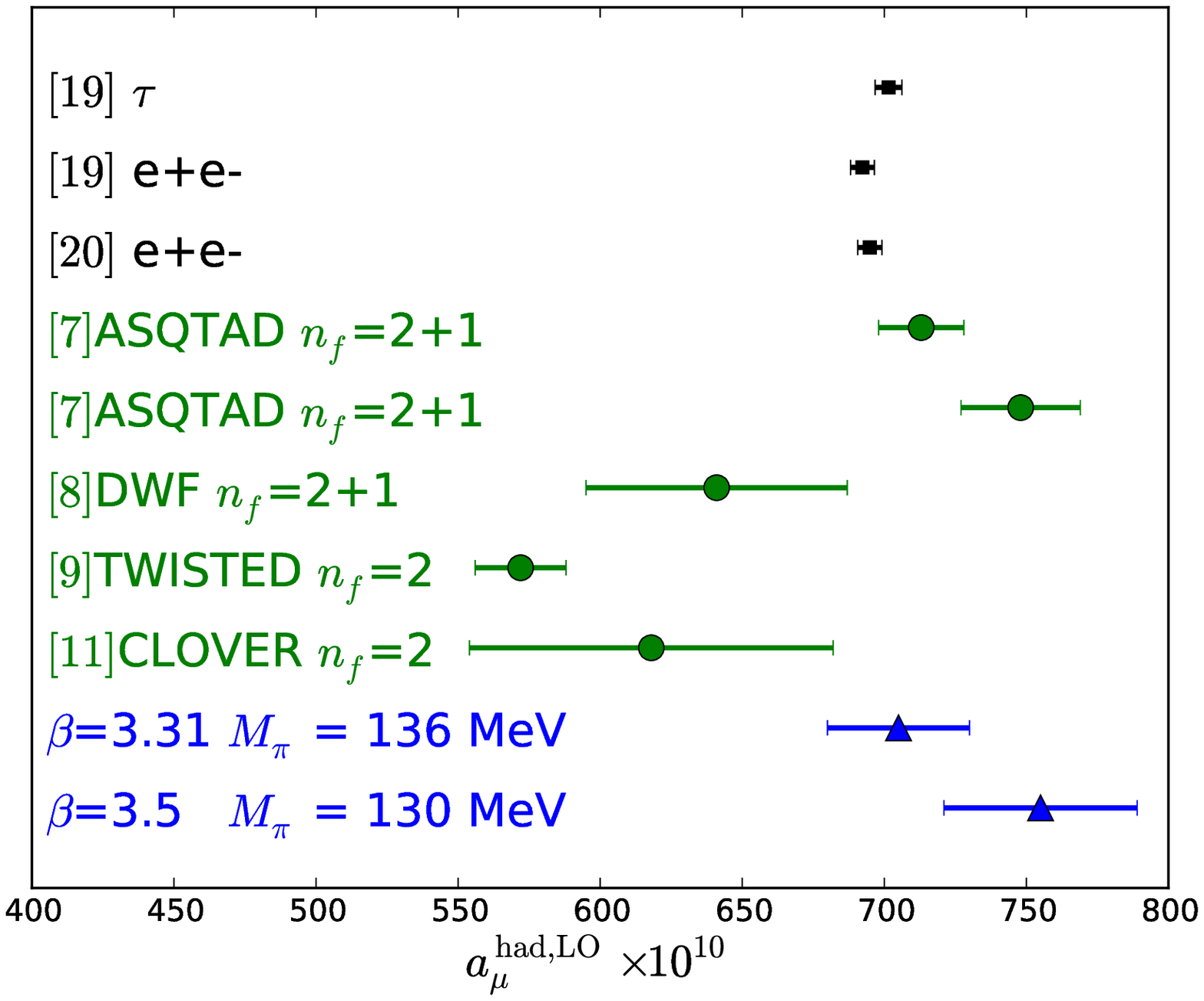}
\end{center}
\vspace{-0.2in}
\caption{Figure $a$ (left)  our results for 
$a_\mu^{\rm had, LO}$ obtained from the various ensembles, as a function of 
$M_\pi^2$, together with those from other groups \cite{Aubin:2006xv, DellaMorte:2012cf, Boyle:2011hu}. In Figure $b$ (right) we compare the 
preliminary results from our two physical $M_\pi$ point simulations (blue triangles)
with those from (black squares) $\tau$ decay and   $e^+e^-$ cross-section data 
\cite{Davier:2010nc,Hagiwara:2011af} and from other lattice computations  (green circles): \cite{Aubin:2006xv,Boyle:2011hu,Feng:2011ff,DellaMorte:2012cf}.
\label{res_summ}}
\end{figure}

\vspace{-0.1in}
In Figs.~\ref{res_summ}a and \ref{res_summ}b we display our preliminary 
results with statistical error bars only. Fig. \ref{res_summ}a shows the value
of $a_\mu^{\rm had, LO}$ we obtain for the various ensembles, as a function of 
$M_\pi^2$. We show results from some other groups for comparison. 
Figure \ref{res_summ}b shows only our physical $M_\pi$ ensemble results 
with other determinations (including calculations with experimental input).

Our future work will refine these calculations, with more ensembles, higher 
statistics and a full error budget. We also plan to include an estimate of the 
disconnected contribution.
\vspace{-0.15in}
\section{Acknowledgments}
\vspace{-0.1in}
The authors thank
%gratefully acknowledge 
the Gauss Centre for Supercomputing (GCS) 
for providing computing time through the John von Neumann Institute for 
Computing (NIC) on the GCS share of the supercomputer JUQUEEN at J\"ulich 
Supercomputing Centre (JSC) and time granted on %the supercomputer 
JUROPA at JSC.
Computations were also performed using HPC resources provided by
GENCI-[IDRIS] (grant 52275).
This work was supported in part by the OCEVU
Labex (ANR-11-LABX-0060) and the A*MIDEX project
(ANR-11-IDEX-0001-02) funded by the ``Investissements
d'Avenir'' French government program managed by the
ANR.  This
work was in part funded by the ``Deutsche Forschungsgemeinschaft'' 
under the grant SFB-TR55.

\end{document}